\documentclass{cflowproc}

\usepackage{natbib}

\begin{document}


\shortauthors{En{\ss}lin, Vogt, \& Pfrommer}     
\shorttitle{Magnetic fields and cosmic rays in cooling flows} 

\title{Magnetic fields and cosmic rays in galaxy cluster cooling flows}   

\author{Torsten En{\ss}lin, 
Corina Vogt, 
and Christoph Pfrommer} 

\affil{}{Max-Planck-Institut f{\"u}r Astrophysik,
  Karl-Schwarzschild-Str. 1, 85741 Garching, Germany} 


\begin{abstract}
Cooling flows are regions where the importance of non-thermal
intra-cluster medium components such as magnetic fields and cosmic
rays may be strongest within a galaxy cluster. They are also regions
where such components are best detectable due to the high gas density
which influences Faraday rotation measurements of magnetic fields and
secondary particle production in hadronic interactions of cosmic ray
nuclei with the ambient thermal gas. New estimates of magnetic fields
in cooling flow and non-cooling flow clusters are presented, which are
based on a newly developed Fourier analysis of extended Faraday
rotation maps. We further present new constraints on the cluster
cosmic ray proton population using radio and gamma-ray observations
measurements of cluster cooling flows, which are especially suited for
this purpose due to their high gas and magnetic energy densities. We
argue that radio synchrotron emission of cosmic ray electrons
generated hadronically by cosmic ray protons is a very plausible
explanation for the radio mini-halos observed in some cooling flows.
\end{abstract}



\section{Cooling flows as a window to non-thermal intra-cluster medium
  components}
\label{Ensslin:intro}

Cooling flows (CFs) are especially well suited places to find traces
of otherwise nearly invisible non-thermal components of the
intra-cluster medium (ICM) due to the extreme gas densities observed
in CF regions. The faded and therefore invisible remnants of radio
galaxy cocoons, so-called {\it radio ghosts}
\citep{1999dtrp.conf..275E} or {\it ghost cavities}, were first
detected in CFs by the absence of X-ray emissivity in the ghost's
volume in contrast to the highly X-ray luminous cooling flow gas
surrounding it \citep[][ and many recent Chandra
observations]{1993MNRAS.264L..25B, 2000MNRAS.318L..65F}. ICM magnetic
fields reveal their presence prominently in CFs by extreme Faraday
rotation measures. Cosmic ray electrons (CRe) are seen in CFs by their
radio synchrotron radiation in the strong CF magnetic fields. Cosmic
ray protons (CRp) in the ICM are most likely to be detected for the
first time within CFs via their hadronic interaction with the dense CF
gas leading to gamma rays and CRe.

A better knowledge of these non-thermal components of the ICM --
especially in the CF regions -- is highly desirable, since they play
important roles in the heat balance of the gas through heating by CRp,
radio ghost buoyant movements, and suppression of heat conduction by
magnetic fields. Additionally, such non-thermal components are tracers
of the violent dynamics of the ICM and may help to solve some of the
puzzles about CFs.

In this article, we present our recent progress in measuring cluster
magnetic fields (Sect.~\ref{Ensslin:RMMF}) and cosmic ray protons
(Sect.~\ref{Ensslin:CRp}) in cooling flows. We focus on ideas and
results, leaving the technical details to the publications given in
the reference list.
  
\section{Faraday rotation \& magnetic fields}
\label{Ensslin:RMMF}

\subsection{Source intrinsic or external?}
\label{Ensslin:external}

\begin{figure}
\plotone{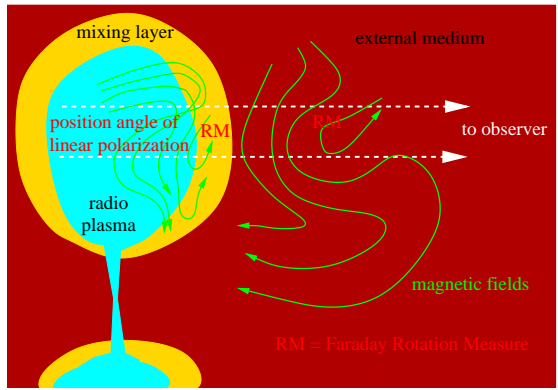} \figcaption{Faraday rotation might be
generated by source intrinsic or extrinsic magnetic fields. In the
first case co-spatial structures in RM and PA are possible, since the
RM would be produced in a mixing layer in which the fields carry
information about the source intrinsic magnetic field geometry, which
determines the PA.
\label{Ensslin:RMPA}}
\end{figure}

The Faraday rotation effect arises whenever polarised radio waves
traverse the magnetised ICM which leads to a rotation of the position
angle of polarisation. This effect allows the measurement of the
(electron density weighted) line of sight component of cluster
magnetic fields. Due to projection, it is not clear where between the
radio emitting volume (typically a radio galaxy cocoon) and the
observer the Faraday rotation occurs (see Fig.~\ref{Ensslin:RMPA}).

In order to test if the Faraday rotation is produced in a mixing layer
between the radio galaxy cocoon and its surrounding ICM,
\citet{2003ApJ...588..143R} compared rotation measure (RM) maps with
intrinsic polarisation position angle (PA) maps of the same radio
source. The PA is parallel to the source intrinsic magnetic fields and
should be uncorrelated to the RM if the Faraday rotation is produced
by a source external magnetic field. Therefore, any real co-spatial
structures in RM and PA maps would be a very strong indicator of an RM
originating from a mixing layer, since only then an RM signature
correlated to the PA could be produced.  \citet{2003ApJ...588..143R}
tested for such co-spatial structures by measuring the clustering in
RM-PA scatter plots and claimed to have found a significant
correlation in the RM and PA maps of a radio galaxy.

However, it was shown by \citet{astro-ph/0301552} that strong
clustering in RM-PA scatter plots appears whenever the RM and PA maps
have patchy structures independently if the maps are correlated or
fully uncorrelated. Such patchy structures are also observed for the
maps analysed by \citet{2003ApJ...588..143R}. In order to test without
bias for potential co-spatial RM and PA structures,
\citet{astro-ph/0301552} developed a method which relies on the
alignment of gradients in the PA and RM maps in a sample of radio
galaxy datasets. No large-scale gradient alignments could be measured,
however a small-scale gradient alignment is present in all maps. Using
a suitable additional test, it could be shown that this alignment
signal is a residual of correlated imperfections arising in the RM and
PA map making process. Such imperfections produce anti-parallel RM and
PA gradients -- as found to be responsible for the full small-scale
alignment signal -- whereas real co-spatial structures should produce
parallel and anti-parallel gradients with the same probability.

Therefore, currently no evidence exists for Faraday rotation being
local to the source. Independently, there is evidence for relatively
strong external magnetic fields by diffuse magneto-synchrotron
emission from galaxy clusters observed as {\it radio halos} and {\it
radio relics}. Furthermore, an RM dispersion excess is
detected for background radio sources seen through galaxy clusters
when compared to control samples \citep[][ and these
proceedings]{2001ApJ...547L.111C}. Additionally, given the
Laing-Garrington effect \citep{1988Natur.331..149L,
1988Natur.331..147G}, we conclude that the assumption of RM being
produced in the foreground ICM of cluster radio sources is strongly
favoured. We will adopt this point of view in the following of this
article.

\subsection{Statistical analysis of RM maps}
\label{Ensslin:RMstat}

\begin{figure}[t]
\plotone{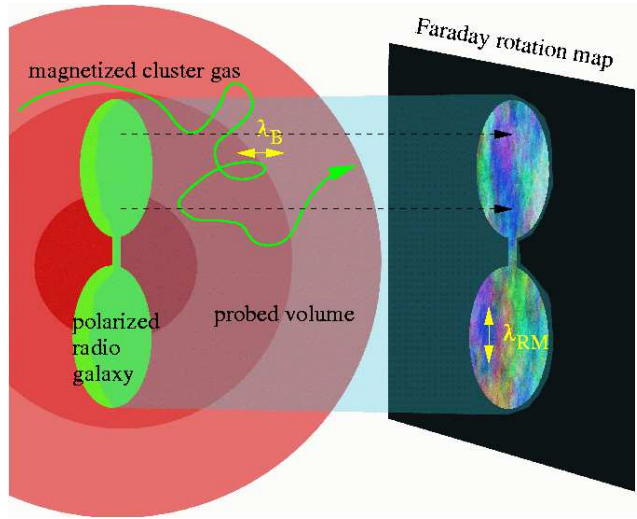} \figcaption{Geometry of a Faraday rotation
measurement of intra-cluster magnetic fields. The volume probed for
magnetic fields is described by a sampling window function which is
non-zero only within the volume in front of the used polarised radio
cocoon. The window function values scale with the cluster gas density,
modelling the structered sensitivity of the Faraday effect to ICM
magnetic fields.
\label{Ensslin:window}}
\end{figure}

A sketch of the typical geometry in which magnetic fields are measured
from Faraday rotation maps is given in Fig.~\ref{Ensslin:window}. The
cluster magnetic field is sampled in a volume - the sampling window
function (or window for short), which is shaped by the geometry of the
used radio source and the gas distribution of the cluster observed
\citep{2003A&A...401..835E}. If the window is sufficiently extended,
the ICM magnetic field can be regarded to be sampled
statistically. This may allow to use statistical methods to restore
some of the information lost by the projection of the fields into the
RM map.

A crucial information for the measurement of magnetic fields from RM
maps is the magnetic autocorrelation length $\lambda_B$. If this is
known, the central magnetic field strength of the cluster would simply
follow from
\begin{equation}
\label{Ensslin:bzlb}
\langle B^2 \rangle = \frac {2\,\langle {\rm RM}^2 \rangle}{a_0
^2\, n_{e} ^2\, L\, \lambda _B} ,
\end{equation}
where $\langle {\rm RM} ^2 \rangle$ is the RM dispersion, $L$ is the
depth of the Faraday screen, which is well defined in terms of the
window function \citep{2003A&A...401..835E}, $a_0\,=\,e^3/(2 \pi
m_e^2c^4)$, and $n_e$ is the (central) electron density.  Here, it is
assumed that the magnetic fields are distributed isotropically, so
that the observed line-of-sight magnetic field component is
representative also for the other two components.

Unfortunately, the magnetic autocorrelation length $\lambda_B$ is not
directly accessible. However, the RM autocorrelation length
$\lambda_{\rm RM}$ of the RM maps is easily measurable. Although these
length scales are often assumed to be equal in the literature, they
are two distinct quantities which usually differ. Under typical
circumstances $\lambda_{\rm RM}$ is expected to exceed
$\lambda_{B}$. The reason for this is that RM maps are more sensitive
to large-scale magnetic structures than to small-scale structures,
since the latter's imprint on the RM maps suffers from stronger
cancelling effects due to field reversals along the
line-of-sight. Therefore, inserting $\lambda_{\rm RM}$ instead of
$\lambda_B$ into Eq.~\ref{Ensslin:bzlb} should lead to an
underestimation of the real magnetic field strength.

An approach to circumvent these difficulties is derived by
\citet{2003A&A...401..835E}: under the assumption of a sufficiently
large window, and statistically isotropic magnetic fields with
vanishing divergence, it can be shown that the 2-dimensional
power-spectrum of RM maps is identical to the 3-dimensional magnetic
field power-spectrum up to a window function dependent (and therefore
known) constant of proportionality. This allows the measurement of the
magnetic power-spectrum $\varepsilon_B(k)$ by Fourier transforming RM
maps. From the magnetic power spectrum, $\lambda_B$ and the average
(cluster central) magnetic field energy density can be calculated.

\subsection{Confronting theory with data}
\label{Ensslin:Hydra}

\begin{figure}[t]
\plotone{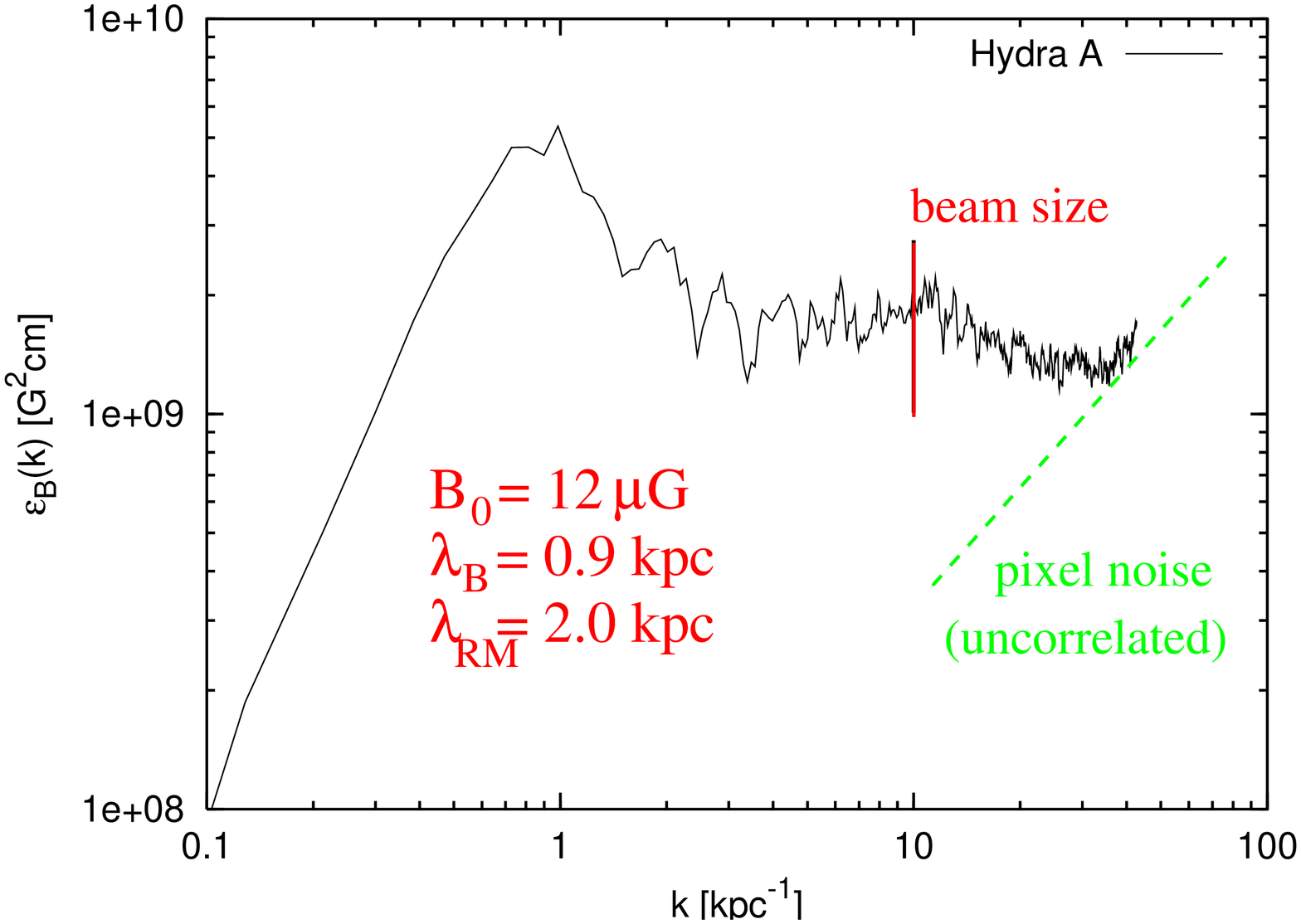} \figcaption{1-dimensional magnetic power
spectrum of magnetic fields in the central region of the Hydra
cluster. The observational beam size and the spectral slope of
uncorrelated sub-beam noise are marked. The estimates of magnetic
field strength and length scales are derived by integrating up to the
$k$-vector corresponding to the beam size.
\label{Ensslin:hydra}}
\end{figure}

The theoretical formalism of \citet{2003A&A...401..835E} was applied
to an RM map of the north lobe of Hydra A by \citet{astro-ph/0309441}.
This polarised radio source is embedded in the Hydra cluster cooling
flow region. The window function was modelled using up-to-date
electron density profiles of the Hydra cluster from X-ray
observations. It was assumed that the northern radio cocoon points
towards the observer with an angle of 45$^\circ$, consistent with the
Laing-Garrington effect observed for this source
\citep{1993ApJ...416..554T}. The resulting power-spectrum is shown in
Fig.~\ref{Ensslin:hydra}. The cluster central magnetic fields are
estimated to have a typical field strength of 12$\,\mu$G, if one
integrates this power spectrum up to the Fourier $k$-vector
corresponding to the observational resolution (beam size). As
predicted, the magnetic autocorrelation length is significantly
shorter than the RM autocorrelation length: $\lambda_B = 0.9$ kpc
compared to $\lambda_{\rm RM} = 2$ kpc. If $\lambda_{\rm RM}$ had been
used instead of $\lambda_B$ the typical magnetic energy density would
have been underestimated by more than a factor of two.

One is tempted to measure the spectral slope of the magnetic power
spectrum from Fig.~\ref{Ensslin:hydra} in order to compare it to
predictions of magneto-hydrodynamical turbulence. Unfortunately, this
is not straightforwardly possible. The difficulties are caused by the
finite size of the window through which the magnetic fields are
observed. Any finite window in real space leads to a redistribution of
power in the Fourier-space. Due to the small size of the window and
its complicated shape, this effect is significant and has to be taken
into consideration. A possibility to investigate this effect is to
calculate theoretically the expected response of the measured power
spectrum to single-scale (isotropic) magnetic fluctuations seen
through the window. 

\begin{figure}[t]
\plotone{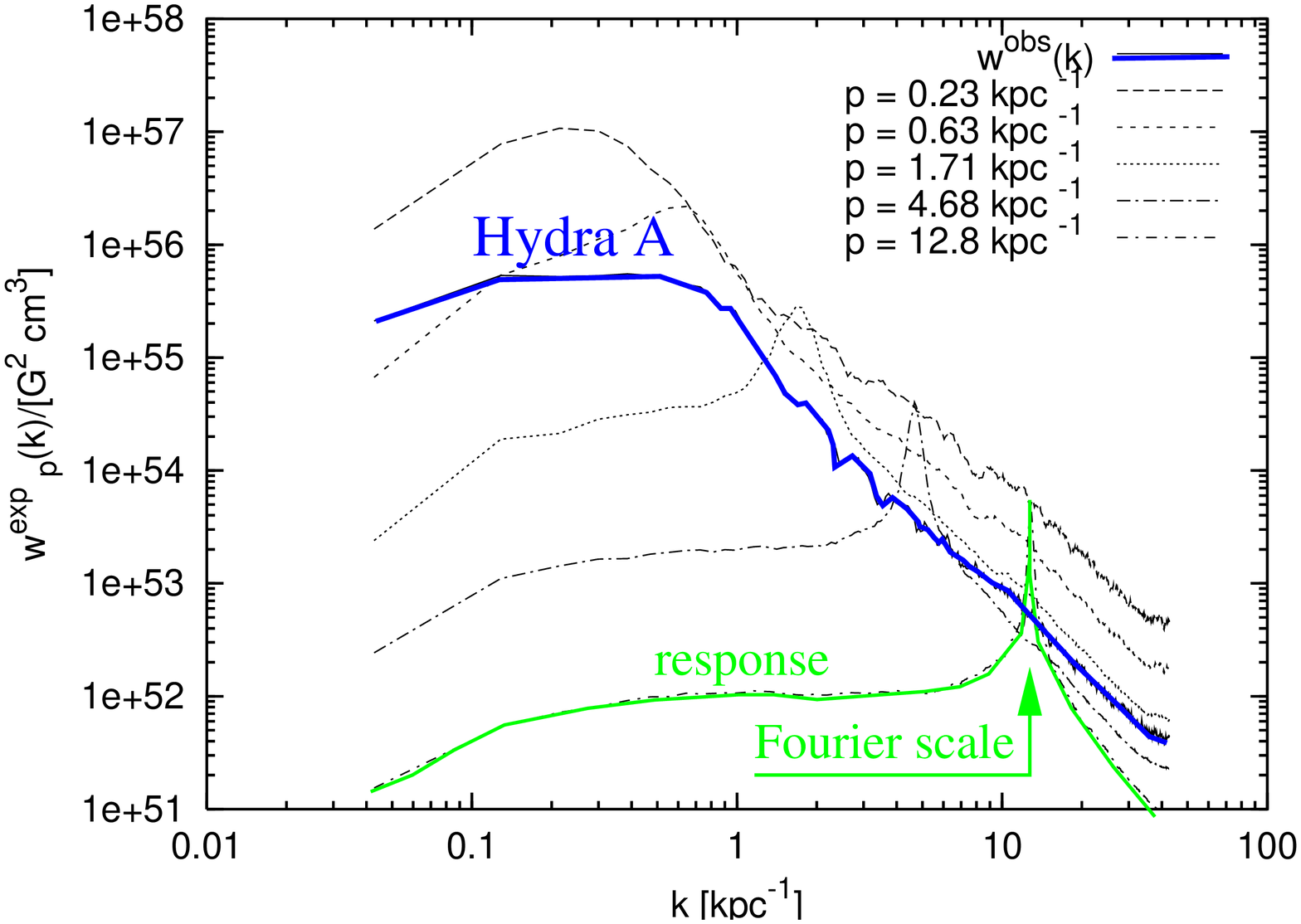} \figcaption{Response to single-scale magnetic
power spectra. The 3-dimensional magnetic power spectrum
$w(k)$ of Hydra A (blue) and the responses of single-scale magnetic
power spectra (delta functions located at Fourier-scale $p$) seen
through the window function of Hydra A are shown.
\label{Ensslin:hydraResponse}}
\end{figure}

The results of such calculations are displayed in
Fig.~\ref{Ensslin:hydraResponse}. The response to magnetic power on
large Fourier-scales, and therefore to small-scale spatial power,
resembles the delta-function. However, magnetic large scale-power is
strongly redistributed over the full Fourier-space. Since the
large-scale power dominates the total power, the intrinsic spectrum of
the magnetic fields in the Hydra A cluster are masked by this
effect. Experiments with power-law spectra observed through the window
reveale that the observational data is consistent with
Kolmogorov-like or steeper spectra.

The redistribution of power in Fourier space affects the accuracy of
magnetic field estimates. Since the estimated magnetic field strengths
are an integral quantity of the power-spectra, the redistribution of
power within the Fourier space typically introduces only a moderate
error. Magnetic field estimates seem to be robust to this effect, at
least on the few 10 \% accuracy level.

\subsection{Improving data: Pacman}
\label{Ensslin:pacman}
Usually, the RM is determined using a least square fit in order to
solve $\Phi={\rm RM}\,\lambda ^2+{\rm PA}$, where $\Phi$ is the measured
position angle of polarisation at the observed wavelength
$\lambda$. However, $\Phi$ is observationally only constrained to
values between 0 and $\pi$ leaving the freedom of additions of $\pm
n\pi$ (where $n$ is an integer) causing the so called $n\pi$
ambiguity. Therefore, the least square fit has to be applied to all
possible $n\pi$-combinations while searching for the
$n\pi$-combination for which $\chi^2$ is minimal.

Traditionally, the minimisation of $\chi^2$ is done for each data
pixel separately. In these RM maps, steps in the RM value from one
pixel to the other could be observed. These steps are mostly detected
for noisy regions. They seem to be artificial and due to a false
solution of the $n\pi$-ambiguity. \citet{2003A&A...401..835E} showed
that artefacts in the RM map can lead to flattening of the power
spectra in their analysis. Thus, magnetic field strengths estimate are
likely to be overestimated.

In order to determine RM values more reliably, Dolag et al. (in prep.)
suggest to use non-local information to solve the
$n\pi$-ambiguity. The new {\it Pacman} algorithm takes advantage of
smooth regions in $\Phi$. For such smooth regions for all observed
wavelengths, the RM distribution is expected to be also smooth and is
not expected to show jump or step like features. Therefore, the
solution to the $n\pi$-ambiguity can be applied to this whole
region. This has the advantage that noisier areas can profit from
leass noisy areas.

The application of {\it Pacman} can lead to a drastic decrease of
noise in RM maps as can be seen in Fig.~\ref{Ensslin:pacman}. In this
figure an RM map which is obtained using standard least square fit
algorithm is shown on the left side. In comparison to it an RM map
generated by {\it Pacman} is exhibited on the right hand side. Already
a visual inspection of this figure reveals that the {\it Pacman} map
is smoother than the standard fit map.

\begin{figure}[t]
\plotone{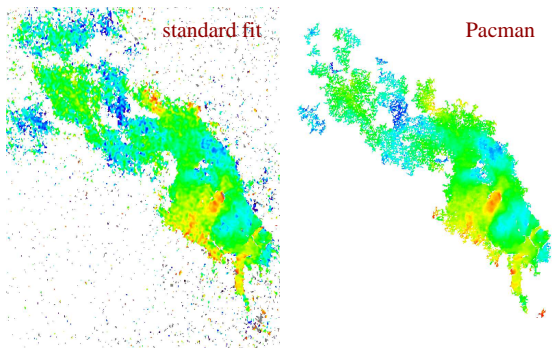} \figcaption{A comparison between an RM map
obtained using standard least square fit routines on the left to an RM
map produced by {\it Pacman} on the right hand side is shown. A
by-eye-comparison reveals that the {\it Pacman} map is smoother. 
\label{Ensslin:pacman}}
\end{figure}

\begin{figure}[t]
\plotone{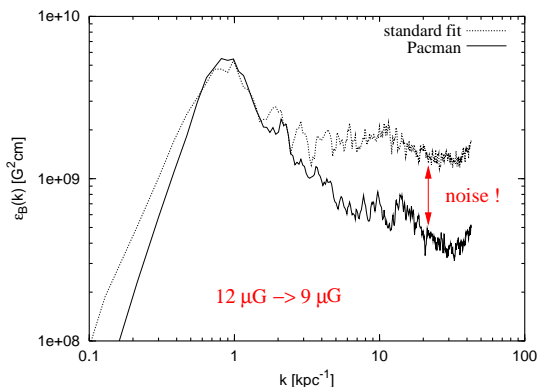} \figcaption{The 1-dimensional power
spectra calculated for the {\it Pacman} map is represented by the
solid line in comparison to the dashed line which was derived by
analysing the standard fit RM map. The influence of the map making
process is clearly visible.
\label{Ensslin:pacman_simp}}
\end{figure}

If one repeats the analysis described above and derives the magnetic
power spectra for the {\it Pacman} map as shown in
Fig.~\ref{Ensslin:pacman_simp}, one can see the effect of noise in the
RM map. The power spectra for the {\it Pacman} map is steeper and
thus, leads to a lower estimate for the central magnetic field
strength of the order of about $\sim$ 9 $\mu$G.

\section{Hadronic interactions of cosmic ray protons}
\label{Ensslin:CRp}
\subsection{Hadronic reactions}
\label{Ensslin:hadronic}
Approximately once in a Hubble time a cosmic ray proton (CRp) collides
inelastically with a nucleon of the ICM gas of non-cooling flow
clusters. Within CFs, such collisions are much more frequent due to
the higher target densities. Such inelastic proton ($p$) nucleon ($N$)
collisions hadronically produce secondary particles like relativistic
electrons, positrons, neutrinos and gamma-rays according to the
following reaction chain:
\begin{eqnarray}
p + N &\rightarrow& 2N + \pi^{\pm/0} 
\nonumber\\
  \pi^\pm &\rightarrow& \mu^\pm + \nu_{\mu}/\bar{\nu}_{\mu} \rightarrow
  e^\pm + \nu_{e}/\bar{\nu}_{e} + \nu_{\mu} + \bar{\nu}_{\mu}\nonumber\\
  \pi^0 &\rightarrow& 2 \gamma \,.\nonumber
\end{eqnarray}
The resulting gamma-rays can be detected directly with current and
future gamma-ray telescopes. The relativistic electrons and positrons
(summarised as CRe) are visible due to two radiation processes:
inverse Compton scattering of background photon fields (mainly the
cosmic microwave background, but also starlight photons) and radio
synchrotron emission in ICM magnetic fields.  Especially the latter
process provides a very sensitive observational signature of the
presence of CRp in cooling flows not only because of the tremendous
collecting area of radio telescopes, but also due to the strong
magnetic fields in CFs, as Faraday rotation measurements demonstrate.

\subsection{Gamma ray constraints from EGRET}
\label{Ensslin:gammas}

\begin{figure}[t]
  \plotone{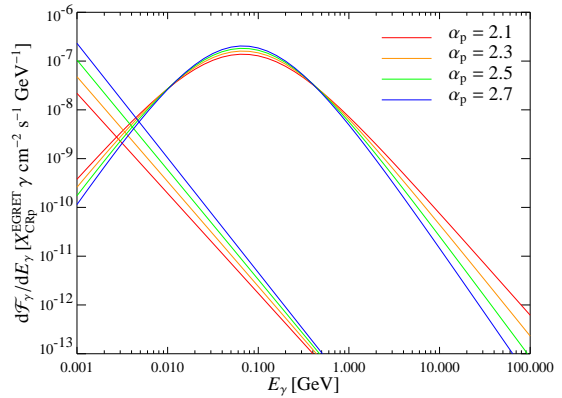} \figcaption{Theoretical gamma-ray spectra of the
    Perseus cluster resulting from CRp hadronic interactions with the CF gas.
    The assumed CRp spectra are normalised to be marginally consistent with the
    EGRET $E_\gamma > 100$ MeV non-detection limits. Results for different
    proton spectral indices are shown. The $\pi^0$ decay bump is clearly
    visible. The low energy part of the spectra is populated by secondary CRe
    inverse Compton scattering cosmic microwave background photons into the
    gamma-ray regime. The CRe population was calculated neglecting synchrotron
    radiation energy losses. Thus, realistic inverse Compton spectra due to
    this process will exhibit a lower normalisation than displayed here.
\label{Ensslin:Perseus}}
\end{figure}

\begin{figure}[t]
  \plotone{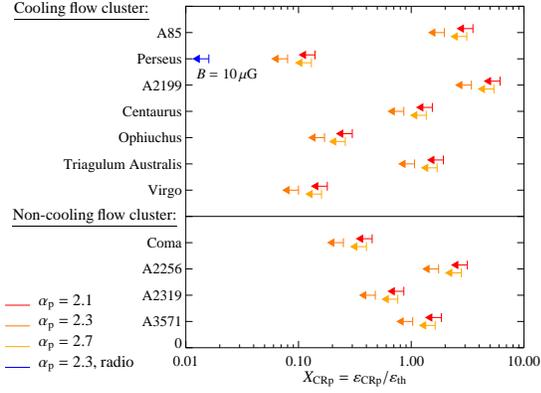} \figcaption{EGRET gamma-ray constraints on
    the (central) CRp energy density ($\varepsilon_{\rm CRp}$, for
    various CRp spectral indices) in terms of the cluster central
    thermal energy density ($\varepsilon_{\rm th}$) for CF and non-CF
    clusters. The constraint from the Perseus cluster radio mini-halo
    observation is also displayed while assuming a typical CF cluster
    magnetic field strength (blue arrow). If radio mini-halos are
    entirely produced hadronically by CRp then the constraint derived
    for the Perseus cluster is an actual measurement for
    $X_\mathrm{CRp}$ and not only an upper limit.
\label{Ensslin:gammaConstr}}
\end{figure}

Assuming that the CRp population can be described by a power law
distribution in momentum, \citet{astro-ph/0306257} developed an
analytical formalism to describe the secondary emission spectra from
hadronic CRp interactions which exhibit the simplicity of textbook
formulae. This formalism was applied to a sample of nearby X-ray
luminous galaxy clusters which are also believed to be powerful gamma
ray emitters owing to the present high target densities. Synthetic
gamma-ray spectra of the Perseus cooling flow cluster calculated using
this formalism are shown in Fig.~\ref{Ensslin:Perseus}. Assuming that
the CRp population follows the spatial distribution of the thermal
ambient intra-cluster gas, we can define the CRp scaling ratio
\begin{equation}
  \label{eq:X_CRp}
  X_\mathrm{CRp} \equiv \frac{\varepsilon_\mathrm{CRp}}{\varepsilon_\mathrm{th}}.
\end{equation}
The parent CRp spectra giving rise to the hadronically induced
gamma-ray spectra are normalised to upper limits on the gamma ray
emission obtained by EGRET observations for energies $E_\gamma > 100$
MeV. For nearby CF clusters, this analysis constrains CRp energy
densities relative to the thermal energy density to $X_\mathrm{CRp}
\sim 10\%$. The real gamma ray spectrum of Perseus is not expected to
be far below these spectra. Many processes like supernova driven
galactic winds, structure formation shock waves, radio galaxy activity
and in-situ turbulent particle acceleration support this
expectation. These processes should have produced a CRp population
characterised by an energy density relative to the thermal energy
density of at least a few percent.

The EGRET gamma ray observation of Perseus limits the CRp content in
the central region to be below $X_\mathrm{CRp} \sim 10\%$. For the
full sample of nearby X-ray luminous clusters upper limits of the same
order of magnitude were obtained as can be seen in
Fig.~\ref{Ensslin:gammaConstr}. It is obvious from this compilation
that CF clusters are extremely well suited to visualise even small CRp
populations.

\subsection{Gamma rays from the central CF region of the Virgo cluster}
\label{Ensslin:virgo}

\begin{figure}[t]
  \plotone{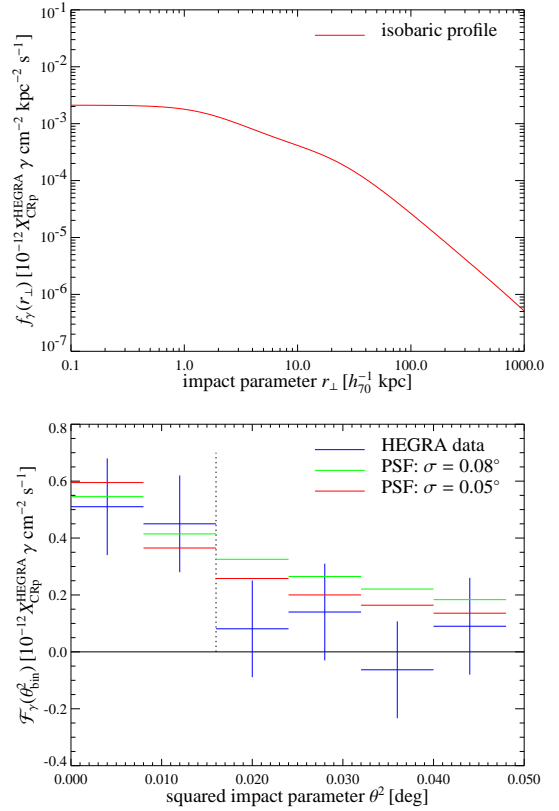} \figcaption{{\bf Upper panel:} Modelled
    $\gamma$-ray surface flux profile of the CF region within the
    Virgo cluster. The profile is normalised by comparing the
    integrated $\gamma$-ray flux above 730~GeV to HEGRA data of M~87
    within the innermost two data points.  {\bf Lower panel:}
    Comparison of detected to predicted $\gamma$-ray flux within the
    central aperture and the innermost annuli for two different widths
    of the PSF. The red lines correspond to an optimistic PSF of
    $\sigma = 0.05\degr$ whereas the green lines are calculated for a
    conservative $\sigma = 0.08\degr$ for a soft gamma ray
    spectrum. The vertical dashed line separates the data from the
    noise level at a position corresponding to $r_\bot = 37.5$~kpc.
\label{Ensslin:Virgo}}
\end{figure}

Recently the HEGRA collaboration \citep{2003A&A...403L...1A} announced
a TeV gamma ray detection from the giant elliptical galaxy M~87 which
is situated at the centre of the CF region of the Virgo cluster.
Using imaging atmospheric \v{C}erenkov techniques, this gamma ray
detection was obtained at a 4-$\sigma$ significance level. On the
basis of their limited event statistics, it is inconclusive whether
the detected emission originates from a point source or an extended
source. Despite testing for eruptive behaviour of M~87, no time
variation of the TeV gamma ray flux has been found.

\citet{2003A&A...407L..73P} applied the previously described
analytical formalism (Sect.~\ref{Ensslin:gammas}) of secondary gamma
ray emission spectra resulting from hadronic CRp interactions to the
central CF region of the Virgo cluster. While combining the observed
TeV gamma ray emission to EGRET upper limits, they constrain the CRp
spectral index $\alpha_\mathrm{GeV}^\mathrm{TeV}<2.3$ provided the
$\gamma$-ray emission is of hadronic origin and the population is
described by a single power-law ranging from the GeV to TeV energy
regime. 

A comparison of the observed to the predicted gamma ray emissivity
profiles is shown in Fig.~\ref{Ensslin:Virgo}. In order to allow for
finite resolution of the \v{C}erenkov telescope, the real profile was
convolved with the point spread function (PSF) of the HEGRA
instrument. The more optimistic PSF of $\sigma = 0.05\degr$ for harder
gamma ray spectra being favoured by our model as well as the
conservative choice of a PSF derived from softer Crab-like spectra are
both consistent with the observed data. When considering an aged CRp
population which has already suffered from significant Coulomb losses,
CRp scaling ratios of the order of $X_\mathrm{CRp} \sim 50\%$ are
obtained for the innermost region of Virgo within $r_\bot = 37.5$~kpc.

Since the emission region is dominated by the giant radio galaxy M~87,
other mechanisms like processed radiation of the relativistic outflow
or dark matter annihilation could also give rise to the observed gamma
ray emission. Nevertheless, the hadronic scenario probes the CRp
population within a mixture of the inter stellar medium of M~87 and
the ICM of the Virgo CF region yielding either an upper limit or a
detection on the CRp population, provided this scenario applies.

\subsection{Radio mini-halos}
\label{Ensslin:mini-halos}

\begin{figure}[t]
  \plotone{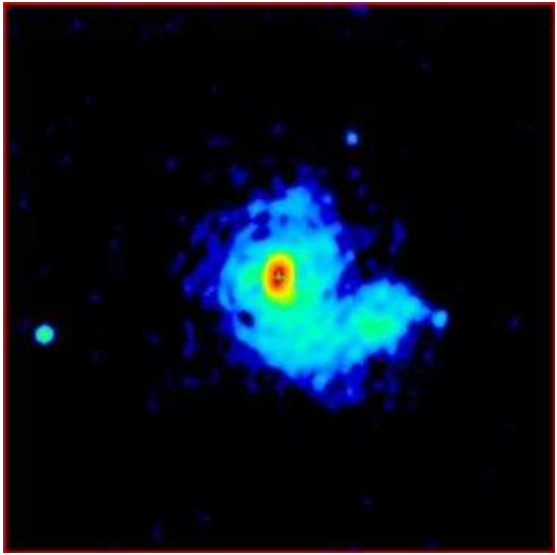} \figcaption{The Perseus radio mini-halo at
1.4 GHz from \citet{1990MNRAS.246..477P} with an extent of $~160
h_{70}^{-1}$~kpc in diameter. At the very centre is an extremely
bright flat-spectrum core owing to relativistic outflows of the radio
galaxy NGC 1275. This outflow was subtracted from this image to make
the extended structure more visible. The position of the radio galaxy
is marked by the green cross in the centre of the red region.
\label{Ensslin:radioPerseuspic}}
\end{figure}

\begin{figure}[t]
  \plotone{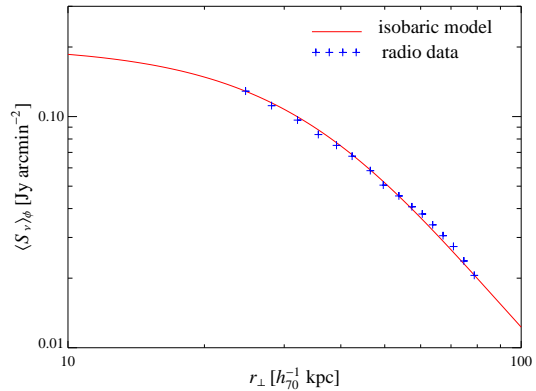} \figcaption{Radio emissivity profile of the
    Perseus mini-halo at 1.4 GHz observed by
    \citet{1990MNRAS.246..477P} (crosses) and predicted due to
    hadronic interactions of CRp by \citet{astro-ph/0306257}. Both
    the required CRp scaling ratio $X_\mathrm{CRp} \simeq 2\%$ and the
    morphological match of the observed and predicted radio
    emissivities strongly indicate the hadronic origin of radio
    mini-halos.
\label{Ensslin:radioPerseus}}
\end{figure}

An upper limit to the CRp population of $X_\mathrm{CRp} \simeq 2\%$ which is
derived from radio observations of the Perseus galaxy cluster is displayed in
Fig.~\ref{Ensslin:gammaConstr}. The radio emissivity of hadronically generated
CRe depends on the assumed magnetic field strength. Thus, upper limits on
$X_\mathrm{CRp}$ rely on the same assumption. However, in the case of strong
magnetic fields (above $3\,\mu$G), the dependence of $X_\mathrm{CRp}$ on the
assumed field strength is very weak, since synchrotron losses dominate in this
regime. If radio mini-halos are entirely produced hadronically by CRp
then the constraint derived for the Perseus cluster is an actual
measurement for $X_\mathrm{CRp}$ and not only an upper limit.

The Perseus radio mini-halo at 1.4~GHz is shown in
Fig.~\ref{Ensslin:radioPerseuspic} as observed by
\citet{1990MNRAS.246..477P}.  The emission due to the relativistic jet
has been subtracted to make the extended structure more visible. The
spatial extent of the radio mini-halo of $~160 h_{70}^{-1}$~kpc in
diameter is too large to be accounted for by synchrotron emission of
direct accelerated CRe in structure formation or accretion
shocks. Thus one needs to consider other injection processes of CRe
into the ICM which are responsible for the observed diffuse radio
emission.  Besides the reacceleration scenario of mildly relativistic
CRe ($\gamma\simeq 100-300$) which are accelerated in-situ by
turbulent Alv\'en waves \citep[][ and these
proceedings]{2002A&A...386..456G}, the hadronic injection scenario of
CRe is a promising alternative.

Azimuthally averaging the radio emission shown in
Fig.~\ref{Ensslin:radioPerseuspic} yields the radio emissivity profile
as displayed in Fig.~\ref{Ensslin:radioPerseus}.  For a comparison, we
overlaid synthetic radio surface brightness profiles resulting from
the hadronic scenario. There is a perfect morphological match between
observed and predicted profiles. Alongside the comparably small
number of CRp required to account for the observed radio mini-halo,
this suggests a hadronic origin of the radio mini-halo in Perseus.

\section{Non-thermal conclusions}
\label{Ensslin:concl}

Our results on magnetic fields and cosmic rays in galaxy cluster
cooling flows can be summarised as follows:\\[1em]
{\bf Magnetic fields in cooling  flows}\\
\begin{itemize}
 \item Sophisticated statistical tests on Faraday rotation maps as
   suggested by \citet{2003ApJ...588..143R} reveal no indications that
   the Faraday effect occurrs in the vicinity of the polarised radio
   source observed \citep{astro-ph/0301552}. Additional several
   independent pieces of evidence in favour of strong intra-cluster
   magnetic fields lead us therefore to believe that the Faraday
   rotation signal is due to intracluster fields, and that the effect
   is not generated in a hypothetical mixing layer surrounding the
   observed radio source.
 \item Faraday rotation measure maps provide a window through which we
   get a glimpse on the turbulent magnetised intra-cluster
   medium. However, they give us only a projected and partial view of
   the cluster magnetic field configuration. Therefore statistical
   methods have to be used to decipher the Faraday signal in terms of
   magnetic field properties.
 \item The magnetic field strength derived from the dispersion of
  rotation measure values depends critically -- besides geometrical
  factors of the galaxy cluster -- on the magnetic autocorrelation
  length, which is not identical to and usually shorter than the
  autocorrelation length of the rotation measure
  fluctuations. However, both can be measured from such maps with a
  novel analysis method \citep{2003A&A...401..835E}.
 \item An application of the novel analysis method of Faraday rotation
  map to data of a Hydra A reveals magnetic fields $\sim 12\,\mu$G in
  the centre of the cooling flow of the Hydra cluster
  \citep{astro-ph/0309441}. The magnetic autocorrelation length is
  0.9~kpc, and therefore shorter than the rotation measure
  autocorrelation length of 2.0~kpc, as expected. The magnetic
  power-spectrum is consistent with a Kolmogorov-like or steeper
  spectrum.
 \item The small-scale fluctuations in the Faraday rotation map of
  Hydra A are dominated by noise. The southern maps reveal several
  step-function-like artefacts which is probably due to the ambiguity
  in the absolute polarisation angle used to determine the rotation
  measure. In order to improve the map quality a new map generating
  algorithm is currently being developed -- {\it Pacman} -- which uses
  non-local information in order to solve angle ambiguities and which
  uses improved fitting methods. A preliminary version of {\it Pacman}
  produces maps with strongly reduced noise level.
\end{itemize}
{\bf Cosmic ray protons in cooling flows}\\
\begin{itemize}
 \item We argue that cooling flows of galaxy clusters are well suited
 to reveal or constrain any cosmic ray proton population via radiation
 from hadronic interactions with the ambient gas nuclei. Such
 collisions lead to gamma-rays and cosmic ray electrons. The former
 would have been seen above 100 MeV by the EGRET telescope if the
 cosmic ray protons had energy densities relative to the thermal gas
 exceeding 10\% \citep{astro-ph/0306257}.
 \item The giant elliptical galaxy M~87 in the centre of the Virgo cluster
 cooling flow region has recently been detected at TeV energies by the
 HEGRA instrument \citep{2003A&A...403L...1A}. These gamma rays could
 be produced by hadronic interactions of a cosmic ray proton
 population if its spectral index
 $\alpha_\mathrm{GeV}^\mathrm{TeV}<2.3$ and its energy density is of
 the order of $50\%$ of the gas within the transition/mixture of
 inter-stellar and intra-cluster medium within M~87
 \citep{2003A&A...407L..73P}.\\
 \item Cosmic ray electrons produced in hadronic interactions are a
very sensitive indicator of cosmic ray protons due to their strong
emissivity. Radio synchrotron emission of such electrons in strong
cooling flow magnetic fields of $\sim 10\,\mu$G limit the cosmic ray
proton energy density to $\sim 2\%$ or less compared to the thermal one in
the Perseus cluster cooling flow \citep{astro-ph/0306257}.
 \item Diffuse radio emission from the Perseus cluster cooling flow
was detected -- the so called Perseus {\it radio mini-halo}. This
radio synchrotron emission may be induced by CRp interactions in the
ICM. This scenario is strongly supported by the very moderate energy
requirements (2\% of the thermal energy) and the excellent agreement
between the observed and the theoretically predicted radio surface
profile \citep{astro-ph/0306257}.
\end{itemize}




\acknowledgements \acknowledgements We thank Tracy Clarke, Klaus
Dolag, Greg Taylor, Francesco Miniati, and Sebastian Heinz for
discussion and collaboration, Alan Pedlar for the permission to
reproduce the Perseus mini-halo image, and the conference organisers
for an excellent meeting.



\bibliography{ensslin}

\begin{thebibliography}{16}
\expandafter\ifx\csname natexlab\endcsname\relax\def\natexlab#1{#1}\fi

\bibitem[{{Aharonian} {et~al.}(2003){Aharonian}, {Akhperjanian}, \&
  {Beilicke}}]{2003A&A...403L...1A}
{Aharonian}, F., {Akhperjanian}, A., \& {Beilicke}, M.~{\it et~al.}. 2003,
  \aap, 403, L1

\bibitem[{{B{\"o}hringer} {et~al.}(1993){B{\"o}hringer}, {Voges}, {Fabian},
  {Edge}, \& {Neumann}}]{1993MNRAS.264L..25B}
{B{\"o}hringer}, H., {Voges}, W., {Fabian}, A.~C., {Edge}, A.~C., \& {Neumann},
  D.~M. 1993, \mnras, 264, L25

\bibitem[{{Clarke} {et~al.}(2001){Clarke}, {Kronberg}, \& {B{\"
  o}hringer}}]{2001ApJ...547L.111C}
{Clarke}, T.~E., {Kronberg}, P.~P., \& {B{\" o}hringer}, H. 2001, \apjl, 547,
  L111

\bibitem[{{En{\ss}lin}(1999)}]{1999dtrp.conf..275E}
{En{\ss}lin}, T.~A. 1999, in Diffuse Thermal and Relativistic Plasma in Galaxy
  Clusters, 275, astro-ph/9906212

\bibitem[{{En{\ss}lin} \& {Vogt}(2003)}]{2003A&A...401..835E}
{En{\ss}lin}, T.~A. \& {Vogt}, C. 2003, \aap, 401, 835

\bibitem[{{En{\ss}lin} {et~al.}(2003){En{\ss}lin}, {Vogt}, {Clarke}, \&
  {Taylor}}]{astro-ph/0301552}
{En{\ss}lin}, T.~A., {Vogt}, C., {Clarke}, T.~E., \& {Taylor}, G.~B. 2003,
  \apj, in press, astro-ph/0301552

\bibitem[{{Fabian} {et~al.}(2000){Fabian}, {Sanders}, {Ettori}, {Taylor},
  {Allen}, {Crawford}, {Iwasawa}, {Johnstone}, \& {Ogle}}]{2000MNRAS.318L..65F}
{Fabian}, A.~C., {Sanders}, J.~S., {Ettori}, S., {Taylor}, G.~B., {Allen},
  S.~W., {Crawford}, C.~S., {Iwasawa}, K., {Johnstone}, R.~M., \& {Ogle}, P.~M.
  2000, \mnras, 318, L65

\bibitem[{{Garrington} {et~al.}(1988){Garrington}, {Leahy}, {Conway}, \&
  {Laing}}]{1988Natur.331..147G}
{Garrington}, S.~T., {Leahy}, J.~P., {Conway}, R.~G., \& {Laing}, R.~A. 1988,
  \nat, 331, 147

\bibitem[{{Gitti} {et~al.}(2002){Gitti}, {Brunetti}, \&
  {Setti}}]{2002A&A...386..456G}
{Gitti}, M., {Brunetti}, G., \& {Setti}, G. 2002, \aap, 386, 456

\bibitem[{{Laing}(1988)}]{1988Natur.331..149L}
{Laing}, R.~A. 1988, \nat, 331, 149

\bibitem[{{Pedlar} {et~al.}(1990){Pedlar}, {Ghataure}, {Davies}, {Harrison},
  {Perley}, {Crane}, \& {Unger}}]{1990MNRAS.246..477P}
{Pedlar}, A., {Ghataure}, H.~S., {Davies}, R.~D., {Harrison}, B.~A., {Perley},
  R., {Crane}, P.~C., \& {Unger}, S.~W. 1990, \mnras, 246, 477

\bibitem[{{Pfrommer} \& {En{\ss}lin}(2003{\natexlab{a}})}]{astro-ph/0306257}
{Pfrommer}, C. \& {En{\ss}lin}, T.~A. 2003{\natexlab{a}}, \aap, in press,
  astro-ph/0306257

\bibitem[{{Pfrommer} \& {En{\ss}lin}(2003{\natexlab{b}})}]{2003A&A...407L..73P}
---. 2003{\natexlab{b}}, \aap, 407, L73

\bibitem[{{Rudnick} \& {Blundell}(2003)}]{2003ApJ...588..143R}
{Rudnick}, L. \& {Blundell}, K.~M. 2003, \apj, 588, 143

\bibitem[{{Taylor} \& {Perley}(1993)}]{1993ApJ...416..554T}
{Taylor}, G.~B. \& {Perley}, R.~A. 1993, \apj, 416, 554

\bibitem[{{Vogt} \& {En{\ss}lin}(2003)}]{astro-ph/0309441}
{Vogt}, C. \& {En{\ss}lin}, T.~A. 2003, \aap, in press, astro-ph/0309441

\end{thebibliography}
\bibliographystyle{apj}

\end{document}